# A New Approach to Characterize Dynamics of ECG-Derived Skin Nerve Activity via Time-Varying Spectral Analysis

Youngsun Kong, *Member, IEEE*, Farnoush Baghestani, *Student Member, IEEE*, William D'Angelo, I-Ping Chen, Ki H. Chon, *Fellow, IEEE*

***Abstract*—Assessment of the sympathetic nervous system (SNS) is one of the major approaches for studying affective states. Skin nerve activity (SKNA) derived from high-frequency components of electrocardiogram (ECG) signals has been a promising surrogate for assessing the SNS. However, current SKNA analysis tools have shown high variability across study protocols and experiments. Hence, we propose a time-varying spectral approach based on SKNA to assess the SNS with higher sensitivity and reliability. We collected ECG signals at a sampling frequency of 10 kHz from sixteen participants who underwent various SNS stimulations. Our spectral analysis revealed that frequency bands between 150 – 1,000 Hz showed significant increases in power during SNS stimulations. Using this information, we developed a time-varying index of sympathetic function measurement based on SKNA, termed Time-Varying Skin Nerve Activity (TVSKNA). TVSKNA is calculated in three steps: time-frequency decomposition, reconstruction using selected frequency bands, and smoothing. TVSKNA indices exhibited generally higher Youden's J, balanced accuracy, and area under the receiver operating characteristic curve, indicating higher sensitivity. The coefficient of variance was lower with TVSKNA indices for most SNS tasks. TVSKNA can serve as a highly sensitive and reliable marker of quantitative assessment of sympathetic function, especially during emotion and stress.**

***Index Terms*— Autonomic Nervous System, Cognitive Stress, Electrocardiogram, Pain, Power Spectral Density Analysis, Skin Nerve Activity, Stroop Test, Sympathetic Function, Thermal Grill, Valsalva Maneuver, Variable Frequency Complex Demodulation**

This work was supported by the Office of Naval Research N00014-21-1-2255. (*Corresponding author:* youngsun.kong.phd@gmail.com).

Youngsun Kong, Farnoush Baghestani, and Ki H Chon are with the department of Biomedical Engineering, University of Connecticut, Storrs, CT 06269 USA.

William D'Angelo is with the Biomedical Systems Engineering and Evaluation Department, Naval Medical Research Unit Department, San Antonio, TX 78234 USA.

I-Ping Chen is with the department of Oral Health and Diagnostic Sciences, University of Connecticut Health, Farmington, CT, 06032, USA

## I. INTRODUCTION

ASSESSMENT of the sympathetic nervous system (SNS) has been one of the major focuses of affective research due to the SNS's association with emotion and stress [1], [2], [3]. In addition, SNS assessment has proven to be valuable in cardiovascular research due to the SNS's association with certain diseases and pathophysiological conditions, such as kidney dysfunction, hypertension, coronary artery disease, and heart failure [4], [5], [6]. The SNS is a branch of the autonomic nervous system and modulates the autonomic neural activities to respond to a potential threat by increasing energy expenditures and inhibiting digestion [7]. After SNS responses, the parasympathetic nervous system steps in to counterbalance the SNS responses [8]. Many studies have devised experiments to induce activation of the SNS and assess its dynamics using various non-invasive measurements, such as electrodermal activity (EDA) [9] and heart rate variability (HRV) [10], which can provide insight into our understanding of pathophysiological conditions and facilitate the development of proper interventions and treatments. For example, detection of SNS activation associated with stress, emotion, and pain has been investigated with EDA [11], [12], [13].

More recently, it has been found that skin nerve activity (SKNA) can be extracted from electrocardiograms (ECG) recorded at a high sampling frequency ($\geq$ 2 kHz) using conventional ECG recording systems [14]. This technique, called neuECG, extracts a higher frequency range of nerve activity compared to ECG and myopotential signals (< 500 Hz), and has shown comparable dynamics to SKNA obtained from microneurography [15]. These high-frequency components mainly originate from the stellate ganglion, which is a key structure in the SNS, responsible for regulating autonomic functions across multiple regions in the body [16]. It has been shown that injection blocking of the stellate ganglion can alleviate post-traumatic stress disorder and limb pain, as these are associated with heightened SNS activity [17]. Laboratory experiments such as cold pressor testing, which elicits SNS, showed elevation of SKNA signals, thereby providing evidence of SKNA as a noninvasive surrogate marker of SNS [15]. A recent study showed that SKNA outperformed EDA in classification of SNS activities [18]. Not only did SKNA show higher performance metrics, but it also provided activation and deactivation time information, which EDA does not [18].

To date, analysis of SKNA has been performed in the time domain [14], [15], [18], [19], [20]. Typically, integrated SKNA (iSKNA) is derived from ECG using a bandpass filter, rectifier, and integration procedures [15]. The most commonly used range for band-pass filtering is 500–1,000 Hz, designed to remove muscle movement artifacts, which predominantly exhibit dynamics below 400 Hz. Microneurography dynamics are often filtered using highpass filters with a cutoff frequency at 700 Hz [15]. While bandpass filtering of 500–1,000 Hz is most popular for SKNA processing, studies have tested various bandpass filter ranges, including 500–700 Hz, 200–1,000 Hz, 700–2,000 Hz, and 1,700–2,000 Hz, depending on the study protocol and experiment, implying inconsistency and high variability in the signal with time-domain approaches [14], [15], [18], [19], [20]. Additionally, while electromyogram (EMG) processing typically uses a low pass filter cutoff at 400 Hz, there is no clear evidence that muscle noise does not affect frequency components higher than 400 Hz [21]. To address the limited and inconsistent information regarding SKNA and EMG dynamics, and the fact that extraction of SKNA indices associated with its dynamics are all performed in the time domain, the aim of this work is to perform extraction of SKNA dynamics via the frequency domain.

In particular, we aim to characterize time-varying dynamic properties of the SNS activity, as their activation and deactivation patterns are fleeting. Therefore, we aim to compare time-varying analysis of SKNA with time-domain indices (i.e., iSKNA) to examine which approach provides better performance measures.

To this end, we first performed the power spectral density analysis (PSD) to examine which frequency components are present in SKNA signals. Using this knowledge from PSD, we then performed a time-frequency spectral (TFS) analysis and then reconstructed an SKNA signal using only the frequency bands associated with its dynamics. Specifically, we used the variable frequency complex demodulation (VFCDM) method for TFS analysis and reconstructed the filtered SKNA signal using the Hilbert transform [22]. This approach is termed **T**ime-**V**arying **S**kin **N**erve **A**ctivity (TVSKNA). To examine TVSKNA's feasibility and compare its effectiveness against iSKNA in discerning dynamics associated with SNS, we performed experiments in human subjects who performed three different tasks that have been widely used to invoke SNS dynamics.

## II. Methods

### *A. Participants*

We recruited eight males and eight females (20–57 years old) as study participants. The participants underwent three tasks that invoked SNS dynamics, including the Valsalva maneuver (VM) [23], Stroop test (ST) [24], and a thermal grill (TG) test [25], in a random order, while two channels of ECG recordings were collected. Signed informed consent forms were obtained prior to each experiment from all participants.

### *B. Stimuli and materials*

Prior to each task (VM, ST, or TG), participants were asked to relax for two minutes for baseline recordings and hemodynamic stabilization.

1) **Valsalva maneuver (VM)**

Valsalva maneuver (VM) involves a forced expiratory effort against a closed airway [23]. VM has been used to diagnose heart conditions or assess the autonomic nervous system [23]. VM increases the number of spikes per burst of muscle sympathetic nerve activity [26]. Therefore, VM is a good procedure to invoke SNS activity. In our experiment, participants were asked to perform a deep inhalation followed by forceful exhalation for 30 seconds. Each participant performed VM three to four times with an approximately 50s interval between trials.

2) **Stroop test (ST)**

The Stroop test (ST), also known as the Stroop Color and Word Test, was designed to assess the cognitive challenge of processing a specific stimulus that is impeded by the simultaneous processing of another stimulus [24]. Studies have shown that the Stroop test increases SNS activity [27], [28]. Participants were presented with one of the following six words on a smart tablet: "Red," "Blue," "Green," "Yellow," "Purple," or "Black." The text color of the words could differ from the written color name. The background color of the screen was also randomly selected to differ from the text color of the word. Participants were asked to state the font color. For example, if "Blue" was written in red color font on a yellow background, the participant was supposed to say "Red." Participants were asked to subvocalize the word's color instead of saying it aloud to minimize speech-related noise. Each stimulus lasted between 1 and 3 seconds, randomized for each participant, and each test session lasted 2 minutes.

3) **Thermal-grill (TG) test**

Thermal-grill (TG) is the perception of burning heat and pain, a sensation illusion, that arises from placing a hand on top of a grill that has interlaced copper tubes with warm (40–50 °C) and cool water (18 °C) in them [25]. Thermal grills have been utilized to safely induce various intensities of heat pain sensations due to temperature differences between warm and cold water tubes. Studies have shown that the TG procedure induces SNS activity [29], [30]. We prepared three TGs to induce varying levels of pain intensities: TG1, TG2, and TG3. The warm water temperature was individually adjusted to induce pain levels 4–6 out of the 10-point visual analog scale (VAS) for TG2 and VAS pain levels > 7 for TG3. For TG1, only cool temperature pipes were connected to the water. In our experiment, participants placed their left hands on the TG. In order to prevent pain anticipation, the participants wore blindfolds. Participants were asked to place their left hand slightly to their left once the verbal cue 'ready' was spoken. Then, an experimenter adjusted a wheeled table to position the target TG directly below the left hand. Upon the 'go' verbal cue, participants promptly lowered their left hand onto the grill. A randomized sequence of 9 stimuli was given for each subject, with 3 stimuli for each thermal grill, with an interstimulus interval of approximately 40s. After each stimulus, participants rated their pain levels on a 0-10 VAS scale.

4) **Electrocardiogram and skin nerve activity**

We collected ECG signals to extract SKNA signals. ECG signals were collected from two different channels at the same time at a sampling frequency of 10k Hz (Figure 1). Channel 1 ECG was collected from two Ag/AgCl electrodes placed on both inner wrists (Lead I). Channel 2 ECG was collected from two Ag/AgCl electrodes placed on the top left side of the chest and below the right rib cage (Lead III). Another Ag/AgCl electrode was attached under the left rib as a reference for both channels. ECG signals were recorded using BioAmp with a PowerLab device (ADInstrument, Sydney, Australia). Details on deriving SKNA from ECG signals are described below (see Section II-D).

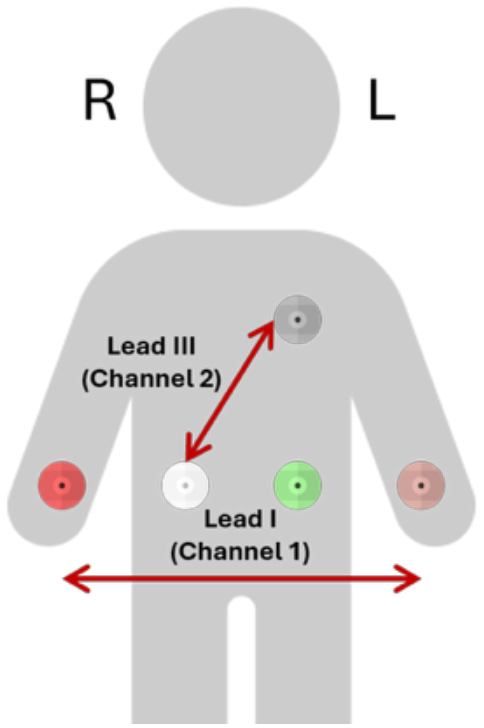

**Fig. 1.** Placement of ECG electrodes.

## C. Design and procedure

Participants were asked to avoid any stimulating or caffeine-containing drink or food starting 24 hours before their experiment day. Before each experiment, participants completed a medical screening questionnaire. The experiments were conducted in a 10 × 10-meter lab in the Engineering and Science building on the Storrs campus of the University of Connecticut. Ag/AgCl electrodes were placed onto participants for ECG recording after cleaning the skin using a cotton ball absorbed with 70% isopropyl alcohol.

Participants then underwent VM, ST, and TG in a random order, with a 2-minute baseline recording for each stimulus task. This research complied with tenets of the Declaration of Helsinki and was approved by the Institutional Review Board at the University of Connecticut.

For TG, two non-sham groups: 1) clinically non-significant pain (CSP-) ($0 < \text{VAS} < 4$) and 2) clinically significant pain (CSP+) ($\text{VAS} \geq 4$) were compared with the baseline. A pain threshold score either less than or greater than 4 is commonly used in clinical settings to categorize between clinically non-significant and significant pain, respectively [31]. Hence, this criterion has been often used for pain quantification studies using physiological signals [32], [33], [34]. We did not analyze the sham group due to uncertainty whether an SKNA increase was due to stress or cold stimulus.

## D. Data processing and analysis

We first performed PSD analysis on ECG signals to identify invoked frequency bands during SNS tasks. Using this information, we decomposed the preprocessed ECG signals into different time-frequency spectral (TFS) components and then reconstructed ECG signals using only those TFS components that represent the invoked SNS frequencies identified by PSD analysis. We named this approach TVSKNA, as defined earlier. We then compared TVSKNA with a time-domain counterpart called iSKNA. However, we first needed to perform preprocessing of ECG to remove non-SKNA frequencies as they would corrupt accurate identification of SKNA dynamics. The next section describes the procedures involved with preprocessing of ECG to remove noise artifacts (non-SKNA dynamics).

1) **Preprocessing to obtain SKNA from ECG**

We down-sampled ECG signals to 4,000 Hz from 10 kHz. A highpass filter with a cutoff frequency of 150 Hz was used for PSD analysis to remove ECG waveforms and powerline interference (e.g., 60 Hz). We identified any noise-related frequencies by visual inspection. Per that inspection, we identified 76 non-SKNA-related frequency regions in both the baseline and the SNS-inducing tasks. These non-SKNA frequencies are likely due to artifact contamination from any of the following sources: bad electrode contact, equipment noise, or environmental interference. We removed these noise sources by applying a series of notch filters using the identified noise frequencies as cutoffs.

2) **Time-varying index**

Once we identified the frequency bands associated with the dynamics of SKNA via PSD, we reconstructed ECG signals using only those frequency components, via a time-varying spectral approach. This process consisted of three steps: 1) decomposition using variable frequency complex demodulation (VFCDM) to obtain the time-frequency spectrum (TFS), 2) instantaneous amplitude estimation using the Hilbert transform, and 3) smoothing with a moving average filter. VFCDM is a technique for time-frequency spectral analysis that exhibits one of the highest TFS resolutions while retaining accurate amplitude estimates when compared to other methods [22]. As VFCDM is detailed in publication [22], we briefly summarize the essence of the method.

First, the ECG signal $x(t)$ was considered as a narrow band oscillation with a center frequency $f_0$, instantaneous amplitude $A(t)$, phase $\varphi(t)$, and the direct current component $dc(t)$, as follows:

$$x(t) = dc(t) + A(t)\cos(2\pi f_0 t + \varphi(t)). \quad (1)$$

For a given center frequency, the instantaneous amplitude information $A(t)$ and phase information $\varphi(t)$ were extracted by multiplying Equation (1) by $e^{-j2\pi f_0 t}$, which results in the following:

$$z(t) = dc(t)e^{-j2\pi f_0 t} + \frac{A(t)}{2} e^{j\varphi(t)} + \frac{A(t)}{2} e^{-j(4\pi f_0 t + \varphi(t))}. \quad (2)$$

By shifting $e^{-j2\pi f_0 t}$ to the left, the center frequency $f_0$ can move to zero frequency in the spectrum of $z(t)$. If $z(t)$ is applied to an ideal low-pass filter (LPF) with a cutoff frequency $f_c < f_0$, then the filtered signal $z_{lp}(t)$ will contain only the component of interest, as follows:

$$z_{lp}(t) = \frac{A(t)}{2} e^{j\varphi(t)}, \quad (3)$$
$$A(t) = 2\left|z_{lp}(t)\right|, \quad (4)$$
$$\varphi(t) = \arctan\left(\frac{imag(z_{lp}(t))}{real(z_{lp}(t))}\right). \quad (5)$$

When the modulating frequency is not fixed but varies as a function of time, the signal $x(t)$ is expressed as follows:

$$x(t) = dc(t) + A(t)\left(\int_0^t \cos(2\pi f(\tau)d\tau + \varphi(t))\right). \quad (6)$$

Similar to Equations (1) and (2), multiplying Equation (6) by $e^{-j\int_0^t 2\pi f(\tau)d\tau}$ produces both instantaneous amplitude, $A(t)$, and instantaneous phase $\varphi(t)$, as follows:

$$z(t) = x(t)e^{-j\int_0^t 2\pi f(\tau)d\tau} = dc(t)e^{-j\int_0^t 2\pi f(\tau)d\tau} + \frac{A(t)}{2} e^{j\varphi(t)} + \frac{A(t)}{2} e^{-j\int_0^t 4\pi f(\tau)d\tau}. \quad (7)$$

From Equation (7), by applying an ideal LPF to $z(t)$ with a cutoff frequency $f_c < f_0$, the filtered signal $z_{lp}(t)$ is obtained with the same instantaneous amplitude $A(t)$ and phase $\varphi(t)$ as provided in Equations (4) and (5). The instantaneous frequency is given by:

$$f(t) = f_0 + \frac{1}{2\pi}\frac{d\varphi(t)}{dt}. \quad (8)$$

VFCDM with a sampling frequency of 4,000 Hz decomposed the signals into centered spectral frequencies ranging from 80 to 1,840 Hz, in 160 Hz increments (a total of 12 components). Figure 2 shows an example of the decomposed signals with their frequency components. We summed the components based on our PSD analysis result (see our Results section). We reconstructed the following summed components to include the sympathetic dynamics: components 2–7 (160–1,120 Hz), components 3–7 (320–1,120 Hz), components 4–7 (480–1,120 Hz), referred to as TVSKNA1, TVSKNA2, and TVSKNA3, respectively, in the rest of the paper. Then, each reconstructed signal was normalized to unit variance. The summed value was denoted by $X'$. Its instantaneous amplitude was then computed using the Hilbert transform, as follows:

$$Y'(t) = \frac{1}{\pi} p.v \int_{-\infty}^{\infty} \frac{X'(\tau)}{t-\tau} d\tau, \quad (9)$$

where $p.v$ represents the Cauchy principal value. As $X'(t)$ and $Y'(t)$ form the complex conjugate pair, an analytic signal, $Z(t)$, is defined as follows:

$$Z(t) = X'(t) + iY'(t) = a(t)e^{j\theta(t)}, \quad (10)$$
$$a(t) = \left[X'^2(t) + Y'^2(t)\right]^{1/2}, \quad (11)$$
$$\theta(t) = \arctan\left(\frac{Y'(t)}{X'(t)}\right). \quad (12)$$

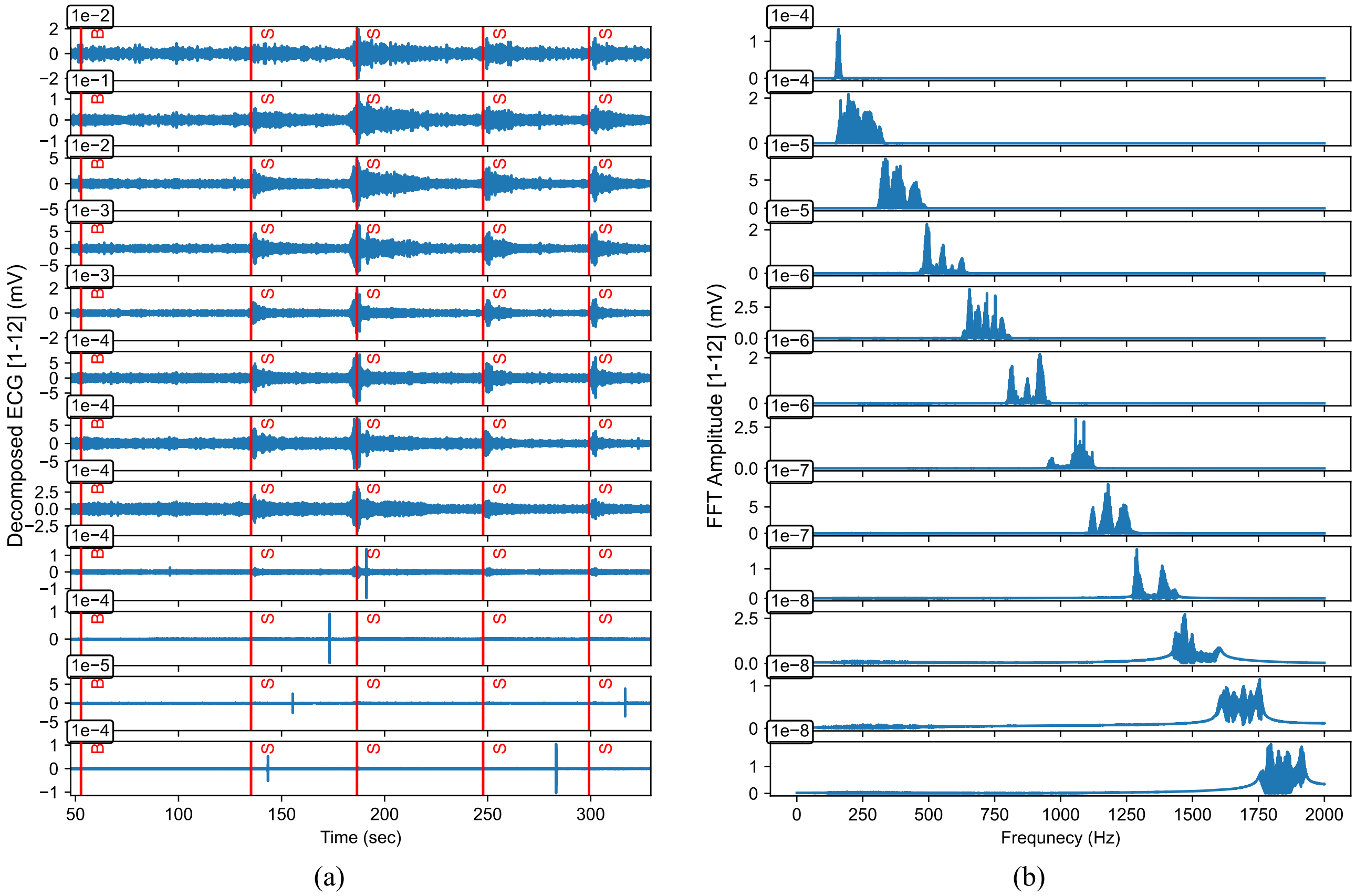

**Fig. 2.** An example of the 12 resulting VFCDM components in (a) time and (b) corresponding frequency domain representation of each time series in (a). Sampling frequency is 4,000 Hz. Red vertical lines indicate the onset of the baseline (B) and the VM stimulation (S) events.

Finally, TVSKNA was calculated by applying a moving average with a 100 ms window to $a(t)$ in Eq. (11). Note that this window duration was previously used in other studies for smoothing [14], [18], [35], [36]. Figure 3 shows an example of TVSKNA computation.

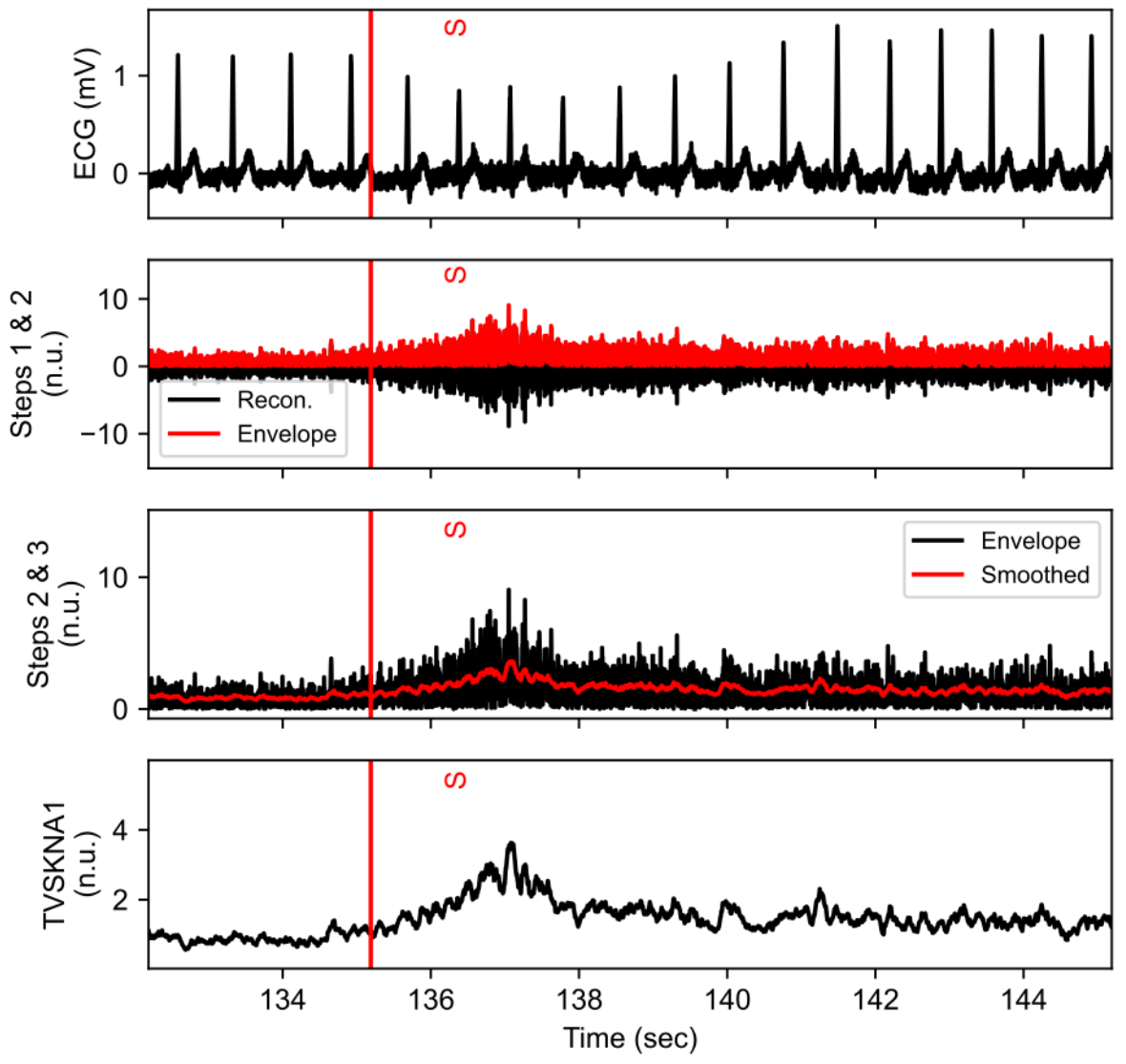

(a)

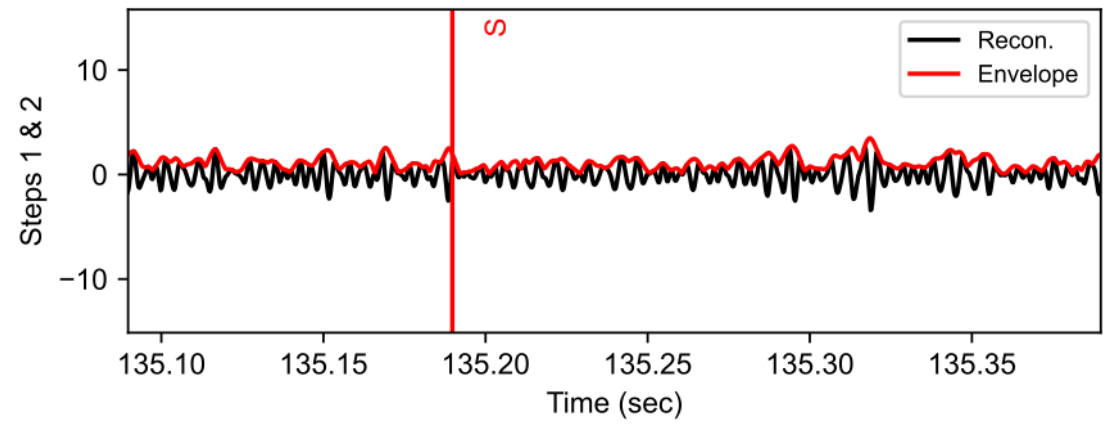

(b)

**Fig. 3.** (a) An example of TVSKNA1 computation (160-1,120 Hz) and (b) zoomed-in plot for Steps 1 and 2. Steps: 1) decomposition (VFCDM), 2) instantaneous amplitude estimation (Hilbert transform), and 3) smoothing (a moving average with 100 ms window). Red vertical lines indicate the onset of the VM stimulation (S) event.

3) **Compared index: time-domain indices of SKNA derived from ECG**

SKNA features are typically extracted between 500 Hz and 1,000 Hz, due to the concern of muscle artifacts. We chose this frequency band since this is most commonly used [14], [15], [18], [19], [20]. We first applied a bandpass filter between 500 Hz and 1,000 Hz. We then rectified the filtered signals and applied a moving average filter with a 100 ms window to obtain iSKNA for each ECG beat. Subsequently, we calculated the maximum, average (aSKNA), and standard deviation (vSKNA) of iSKNA signals. Figure 4 shows an example of iSKNA computation.

4) **Performance metrics and statistics**

To evaluate the performance of each method's (iSKNA and TVSKNA) indices, we extracted the maximum, mean, and standard deviation of each index measured from the baseline and the SNS-inducing task segments. We employed segment sizes of 30s, 120s, and 10s for the VM, ST, and TG tasks, respectively, due to the varying stimulus durations of each task. SKNA indices were then compared between the baseline and each SNS task. To evaluate the performance of SKNA indices, we calculated the optimal (i.e., the highest value) Youden's index (J = Sensitivity + Specificity – 1) [37] and the balanced accuracy (average of sensitivity and specificity), and the area under the curve (AUC) from the receiver operating characteristic (ROC) curves of each experiment task. The ROC curves were obtained for each SKNA index with baseline measurements designated as the negative classes and task measurement data as positive classes. ROC curves calculate true positive and false positive rates of a classification model at all classification thresholds that determine two classes (baseline and task). We used the pROC package in R to calculate ROC curves [38]. Additionally, we calculated the coefficient of variation (the standard deviation of measurements divided by the mean) [39] and the intraclass correlation coefficients (ICC) [40] for all SKNA indices to assess variability and reliability of the index across subjects, respectively. According to Koo and Li's guideline, values less than 0.5, between 0.5 and 0.75, between 0.75 and 0.9, and greater than 0.9 are indicative of poor, moderate, good, and excellent reliability, respectively [40]. We used the psych package in R to calculate ICC [41]. To see whether statistical differences existed between the baseline and each experimental task, we fit linear mixed-effects models using the lmer package in R [42]. We chose this approach to account for various measurement numbers of each subject for VM and TG experiments. We considered *p*-value < .05 to be statistically significant.

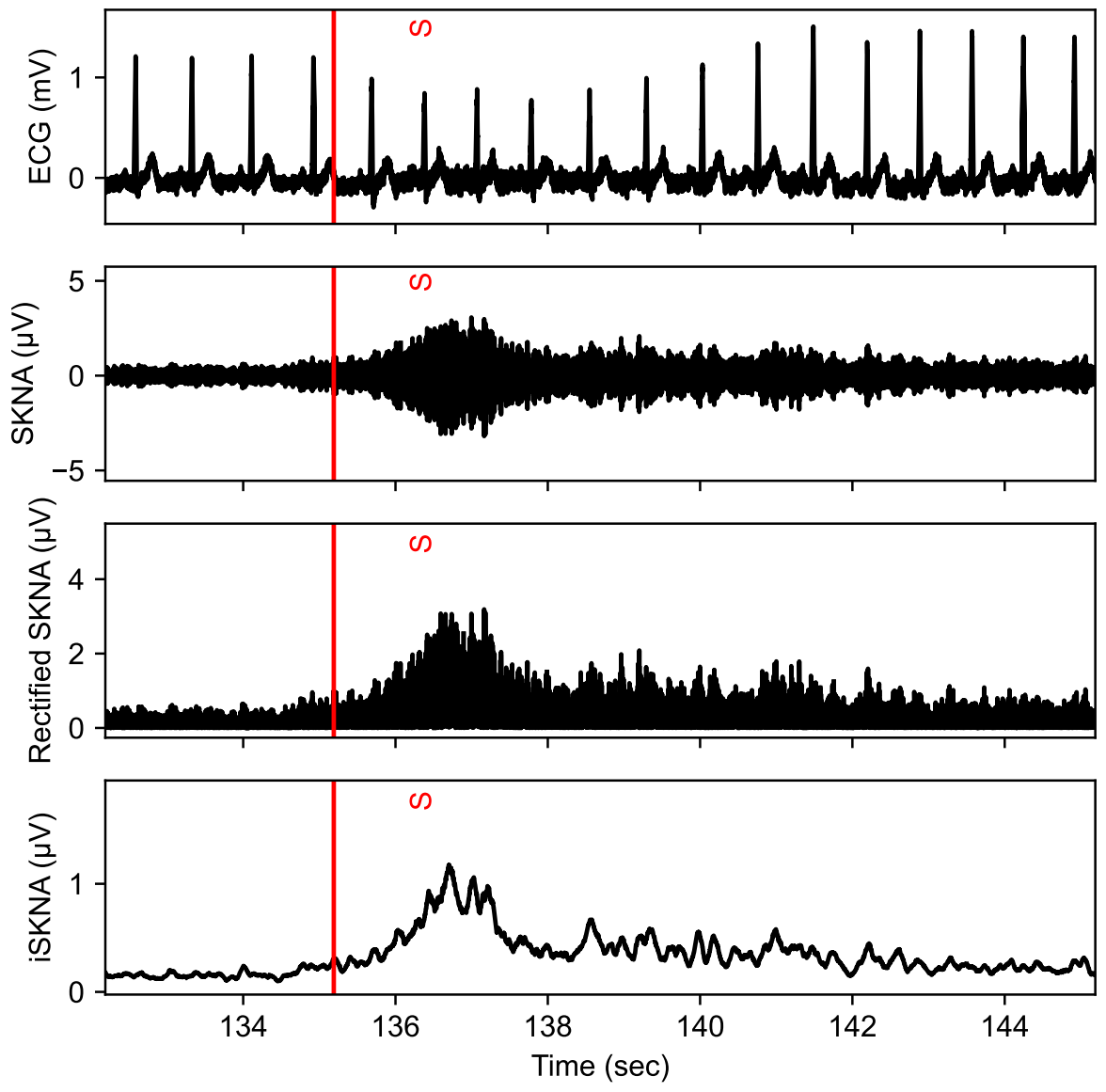

(a)

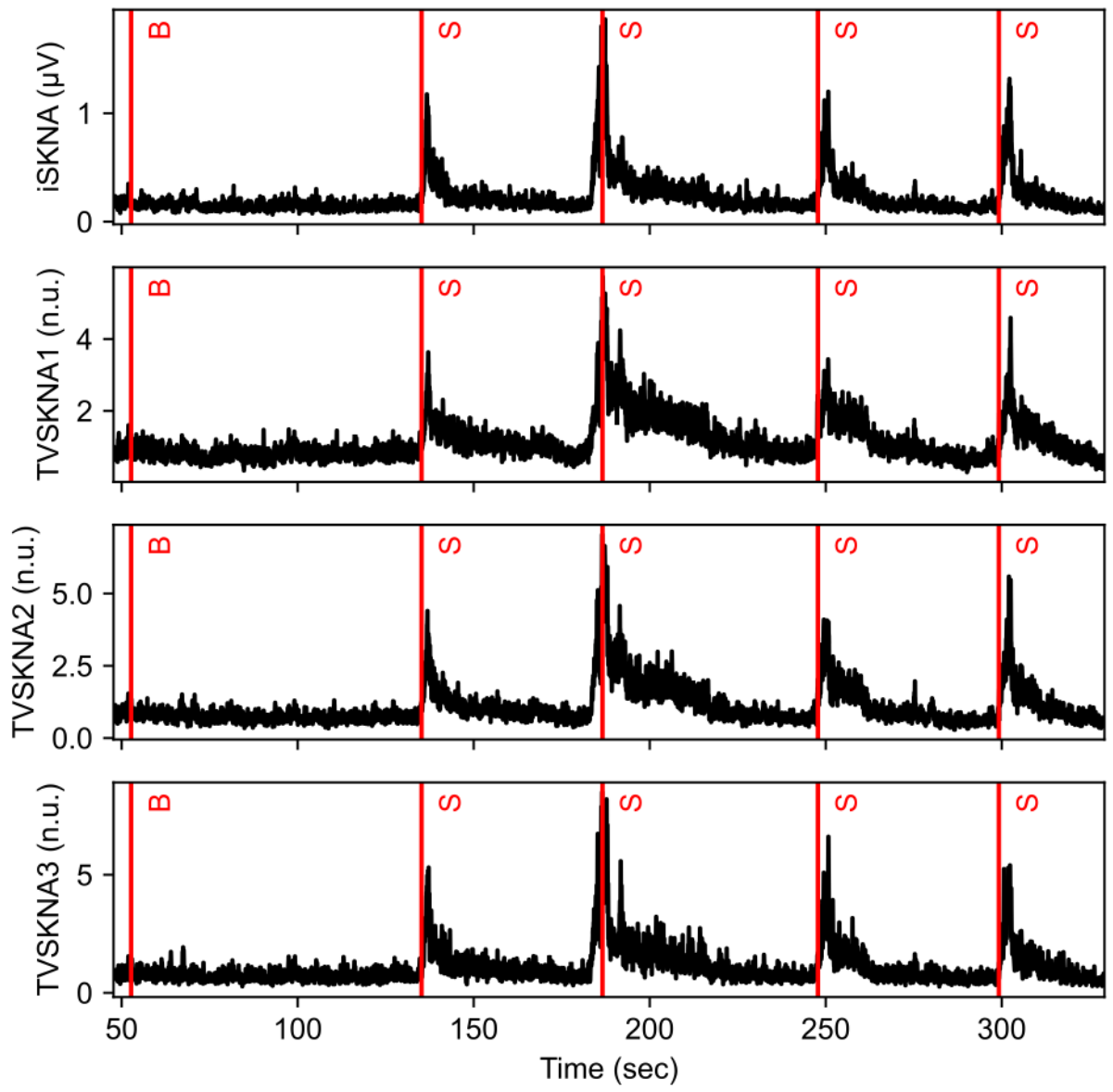

(b)

**Fig. 4.** An example of (a) iSKNA computation and (b) a comparison between iSKNA and TVSKNA. TVSKNA was reconstructed using the following frequency components: 160–1,120 Hz (TVSKNA1), 320–1,120 Hz (TVSKNA2), and 480–1,120 Hz (TVSKNA3). Red vertical lines indicate the onset of the baseline (B) and the VM stimulation (S) events.

## III. Results

A few segments were excluded from this study due to poor data quality, including sensor, noise, and other issues. Table I shows the total number of subjects and segments that were used for each analysis.

TABLE I

The total number of subjects and segments.

| | | Collected | | Available | |
|---|---|---|---|---|---|
| | # of | Subjects | Segments | Subjects | Segments |
| VM | Channel 1 | 16 | 52 | 15 | 47 |
| | Channel 2 | 16 | 52 | 15 | 49 |
| ST | Channel 1 | 16 | 16 | 14 | 14 |
| | Channel 2 | 16 | 16 | 14 | 14 |
| TG | Channel 1 | 16 | 144 | 15 | 135 |
| | Channel 2 | 16 | 144 | 15 | 115 |

VM: Valsalva maneuver, ST: Stroop test, TG: thermal grill

### A. Power spectral density analysis

Table II shows the results of PSD analysis of ECG data, highpass filtered at a cutoff frequency of 150 Hz, for sympathetic function assessment. We focused on the increase in absolute power rather than normalized power change (absolute power divided by total power), because the absolute power can still increase even if the proportional power decreases. For example, in the VM (Channel 1) from 150 to 250 Hz, the normalized power decreased compared to the baseline (63.4% → 57.7%), but the absolute power significantly increased in VM compared to the baseline (indicated by asterisks and superscripts in Table II). All SNS tasks in both ECG channels exhibited the most power within 150–500 Hz. VM (Channel 1) exhibited significant increases in absolute power across all frequency bands (150–2,000 Hz). Both ECG channels exhibited significant increases in absolute power across frequency bands between 150–1,000 Hz during most SNS tasks compared to the baseline. For CSP-, both ECG channels showed significant increases in absolute power across frequency bands between 150–750 Hz compared to the baseline. However, only Channel 1 ECG showed significantly higher absolute power in the 750–1,500 Hz range in CSP+ compared to CSP-. Interestingly, ST (Channel 1) showed significant increases in absolute power across frequency bands ranging from 150–1,250 Hz, whereas ST (Channel 2) exhibited no significant increase for any frequency band.

### B. Time-varying index

Table III shows the maximum, mean, and standard deviation values of iSKNA and TVSKNA indices. For VM (both ECG channels), all indices showed significantly higher values compared to the baseline ($p < .001$). For ST, the maximum and/or mean values of iSKNA and TVSKNA (Channel 1) were significantly higher than the baseline values ($p < .05$), whereas none of the indices (Channel 2) showed significant increases. For the TG pain test, all indices (both Channels) of both CSP- and CSP+ exhibited significantly higher values than those of the baseline.

TABLE II

Power Spectral Density Analysis of SKNA for Sympathetic Function Assessment (mean ± s.d.).

| Channel 1 | VM | | ST | | TG | | |
|---|---|---|---|---|---|---|---|
| Frequency range (Hz) | Baseline (%) | Stimulation (%) | Baseline (%) | Stimulation (%) | Baseline $^{0}$ (%) | CSP-$^{1}$ (%) | CSP+$^{2}$ (%) |
| 150–250 | 63.367±5.134 | 57.73±5.369** | 61.171±4.576 | 63.143±5.087* | 63.749±5.718 | 62.333±4.493$^{0}$ | 61.656±4.171$^{0}$ |
| 250–500 | 36.264±5.113 | 41.826±5.297** | 38.475±4.562 | 36.525±5.062* | 35.902±5.714 | 37.344±4.48$^{0}$ | 37.977±4.122$^{0}$ |
| 500–750 | 0.166±0.049 | 0.275±0.102** | 0.167±0.055 | 0.147±0.053* | 0.141±0.061 | 0.151±0.041$^{0}$ | 0.184±0.086$^{0}$ |
| 750–1,000 | 0.005±0.002 | 0.009±0.004** | 0.005±0.003 | 0.004±0.002* | 0.005±0.003 | 0.004±0.002 | 0.009±0.013$^{0,1}$ |
| 1,000–1,250 | 0.005±0.003 | 0.006±0.002** | 0.006±0.004 | 0.004±0.002* | 0.006±0.003 | 0.004±0.003 | 0.013±0.024$^{0,1}$ |
| 1,250–1,500 | 0.0±0.0 | 0.0±0.0** | 0.0±0.0 | 0.0±0.0* | 0.0±0.0 | 0.0±0.0 | 0.0±0.0$^{0,1}$ |
| 1,500–1,750 | 0.0±0.0 | 0.0±0.0** | 0.0±0.0 | 0.0±0.0* | 0.0±0.0 | 0.0±0.0$^{0}$ | 0.0±0.0$^{0}$ |
| 1,750–2,000 | 0.0±0.0 | 0.0±0.0** | 0.0±0.0 | 0.0±0.0 | 0.0±0.0 | 0.0±0.0$^{0}$ | 0.0±0.0$^{0}$ |
| **Channel 2** | **VM** | | **ST** | | **TG** | | |
| **Frequency range (Hz)** | **Baseline (%)** | **Stimulation (%)** | **Baseline (%)** | **Stimulation (%)** | **Baseline $^{0}$ (%)** | **CSP-$^{1}$ (%)** | **CSP+$^{2}$ (%)** |
| 150–250 | 63.701±8.61 | 55.517±5.189** | 63.231±9.275 | 61.435±7.55 | 63.887±8.737 | 62.218±3.955$^{0}$ | 61.098±5.692$^{0}$ |
| 250–500 | 35.7±8.342 | 43.93±5.039** | 36.277±9.19 | 38.066±7.468 | 35.609±8.659 | 37.402±3.978$^{0}$ | 38.474±5.662$^{0}$ |
| 500–750 | 0.278±0.265 | 0.36±0.149** | 0.241±0.126 | 0.251±0.124 | 0.225±0.133 | 0.163±0.031$^{0}$ | 0.228±0.075$^{0}$ |
| 750–1,000 | 0.027±0.067 | 0.017±0.018** | 0.013±0.013 | 0.014±0.014 | 0.012±0.013 | 0.006±0.002 | 0.011±0.009$^{0}$ |
| 1,000–1,250 | 0.083±0.268 | 0.028±0.064* | 0.017±0.023 | 0.022±0.028 | 0.019±0.023 | 0.005±0.002 | 0.012±0.012 |
| 1,250–1,500 | 0.001±0.003 | 0.0±0.001 | 0.0±0.0 | 0.0±0.0 | 0.0±0.001 | 0.0±0.0 | 0.0±0.0 |
| 1,500–1,750 | 0.0±0.0 | 0.0±0.0* | 0.0±0.0 | 0.0±0.0 | 0.0±0.0 | 0.0±0.0 | 0.0±0.0 |
| 1,750–2,000 | 0.006±0.022 | 0.001±0.004 | 0.0±0.0 | 0.001±0.001 | 0.0±0.0 | 0.0±0.001 | 0.001±0.002 |

Asterisks (* $p < .05$ and ** $p < .001$) indicate a significant increase in absolute power (not normalized power) compared to the baseline. Superscripts (0: Baseline, 1: CSP-, 2: CSP+) for pain analysis indicate significantly higher than the corresponding column ($p < .05$).

TABLE III

RESULTS OF iSKNA AND TVSKNA INDICES (MEAN ± S.D.).

| Channel 1 | | VM | | ST | | TG | | |
|---|---|---|---|---|---|---|---|---|
| | | Baseline | Stimulation | Baseline | Stimulation | Baseline$^{0}$ | CSP-$^{1}$ | CSP+$^{2}$ |
| iSKNA | Max | 0.54±0.21 | 1.97±0.56** | 0.68±0.25 | 0.98±0.6 | 0.39±0.16 | 1.91±0.8$^{0}$ | 3.17±2.04$^{0,1}$ |
| | Mean | 0.22±0.07 | 0.37±0.11** | 0.21±0.06 | 0.26±0.05* | 0.22±0.1 | 0.68±0.23$^{0}$ | 0.82±0.28$^{0}$ |
| | S.D. | 0.06±0.02 | 0.29±0.09** | 0.07±0.03 | 0.08±0.03 | 0.05±0.02 | 0.32±0.16$^{0}$ | 0.43±0.18$^{0,1}$ |
| TVSKNA1 | Max | 1.57±0.41 | 4.69±1.47** | 2.3±0.78 | 3.27±1.43* | 0.9±0.31 | 3.88±0.75$^{0}$ | 5.54±1.87$^{0,1}$ |
| | Mean | 0.86±0.22 | 1.33±0.33** | 0.97±0.2 | 1.3±0.13** | 0.56±0.23 | 1.73±0.22$^{0}$ | 1.9±0.32$^{0}$ |
| 160-1,120 Hz | S.D. | 0.16±0.05 | 0.7±0.27** | 0.25±0.12 | 0.3±0.11 | 0.1±0.03 | 0.64±0.28$^{0}$ | 0.83±0.31$^{0,1}$ |
| TVSKNA2 | Max | 1.74±0.55 | 5.73±1.93** | 2.61±0.99 | 3.66±1.47* | 0.9±0.35 | 4.57±1.06$^{0}$ | 5.74±2.18$^{0,1}$ |
| | Mean | 0.85±0.25 | 1.33±0.34** | 1.04±0.19 | 1.26±0.14* | 0.52±0.21 | 1.81±0.2$^{0}$ | 1.92±0.35$^{0}$ |
| 320-1,120 Hz | S.D. | 0.19±0.06 | 0.86±0.31** | 0.29±0.13 | 0.33±0.12 | 0.1±0.03 | 0.72±0.3$^{0}$ | 0.88±0.34$^{0,1}$ |
| TVSKNA3 | Max | 2.0±0.81 | 7.39±2.82** | 3.36±1.37 | 4.89±2.09* | 0.9±0.28 | 5.33±1.09$^{0}$ | 7.07±4.41$^{0,1}$ |
| | Mean | 0.78±0.26 | 1.32±0.35** | 1.03±0.18 | 1.27±0.15* | 0.44±0.16 | 1.98±0.73$^{0}$ | 1.82±0.68$^{0}$ |
| 480-1,120 Hz | S.D. | 0.23±0.09 | 1.08±0.38** | 0.36±0.13 | 0.43±0.12 | 0.12±0.03 | 0.83±0.18$^{0}$ | 1.01±0.5$^{0}$ |
| Channel 2 | | VM | | ST | | TG | | |
| | | Baseline | Stimulation | Baseline | Stimulation | Baseline$^{0}$ | CSP-$^{1}$ | CSP+$^{2}$ |
| iSKNA | Max | 0.41±0.22 | 1.67±0.66** | 0.47±0.22 | 0.67±0.49 | 0.28±0.09 | 0.4±0.14$^{0}$ | 0.68±0.54$^{0}$ |
| | Mean | 0.16±0.09 | 0.29±0.12** | 0.13±0.04 | 0.15±0.05 | 0.13±0.04 | 0.19±0.06$^{0}$ | 0.22±0.07$^{0}$ |
| | S.D. | 0.05±0.03 | 0.26±0.1** | 0.05±0.02 | 0.05±0.03 | 0.04±0.01 | 0.06±0.03$^{0}$ | 0.09±0.06$^{0}$ |
| TVSKNA1 | Max | 1.82±0.38 | 6.82±2.14** | 3.19±1.02 | 3.63±1.24 | 1.81±0.3 | 3.08±0.64$^{0}$ | 4.19±2.62$^{0}$ |
| | Mean | 0.79±0.19 | 1.37±0.19** | 1.04±0.2 | 1.17±0.16 | 0.86±0.15 | 1.5±0.22$^{0}$ | 1.49±0.28$^{0}$ |
| 160-1,120 Hz | S.D. | 0.26±0.07 | 1.11±0.34** | 0.38±0.1 | 0.39±0.09 | 0.27±0.09 | 0.48±0.21$^{0}$ | 0.57±0.3$^{0}$ |
| TVSKNA2 | Max | 1.82±0.72 | 7.37±1.79** | 3.42±1.37 | 3.92±1.96 | 1.73±0.42 | 3.12±0.93$^{0}$ | 4.24±2.31$^{0}$ |
| | Mean | 0.79±0.2 | 1.39±0.2** | 1.11±0.21 | 1.19±0.18 | 0.85±0.17 | 1.54±0.24$^{0}$ | 1.55±0.32$^{0}$ |
| 320-1,120 Hz | S.D. | 0.23±0.09 | 1.2±0.34** | 0.37±0.13 | 0.37±0.12 | 0.23±0.08 | 0.49±0.27$^{0}$ | 0.58±0.31$^{0}$ |
| TVSKNA3 | Max | 2.14±0.67 | 8.67±2.17** | 4.03±1.45 | 4.92±2.59 | 1.79±0.6 | 4.06±1.63$^{0}$ | 4.35±2.54$^{0}$ |
| | Mean | 0.75±0.21 | 1.39±0.23** | 1.09±0.21 | 1.21±0.17 | 0.75±0.18 | 1.69±0.7$^{0}$ | 1.45±0.54$^{0}$ |
| 480-1,120 Hz | S.D. | 0.27±0.09 | 1.29±0.35** | 0.42±0.13 | 0.44±0.13 | 0.25±0.08 | 0.59±0.2$^{0}$ | 0.61±0.37$^{0}$ |

Asterisks (* $p < .05$ and ** $p < .001$) indicate a significant increase in each index compared to the baseline. Superscripts (0: Baseline, 1: CSP-, 2: CSP+) for pain analysis indicate significantly higher than the corresponding column ($p < .05$).

Table IV shows our sensitivity analysis of iSKNA and TVSKNA indices. For VM, CSP-, and CSP+ (Channel 1), iSKNA Max exhibited the value of 1 for all metrics. For VM Channel 2, the maximum, mean, and standard deviation of TVSKNA indices were higher than or the same as all corresponding iSKNA indices (e.g., iSKNA mean vs. TVSKNA1 mean). For ST, while most indices exhibited lower metric values compared to other tasks for both channels, TVSKNA1 mean showed the highest sensitivity metrics for both channels. For CSP- and CSP+, all TVSKNA indices (Channel 1) showed the value of 1 for all metrics, while TVSKNA1 Max and Mean showed the highest metric values. Additionally, TVSKNA2 showed the highest metric values, with 0.92 for Youden's J, 0.96 for balanced accuracy, and 0.99 for AUC.

Table V shows the average coefficient of variance between the baseline and each SNS task and intra-class correlation results. All indices for VM, CSP-, and CSP+ showed excellent reliability for both channels (ICC ≥ 0.9). ST showed relatively inconsistent results across subjects with lower ICCs; none of the indices during ST in Channel 2 showed either good or excellent reliability (ICC < 0.75). The coefficient of variance was generally lower with TVSKNA indices, except VM Channel 1.

## IV. DISCUSSION

In this work, we proposed a new time-frequency approach to analyze highly sampled ECG to assess SNS response. Our approach is a time-varying spectral approach, whereas the current methods rely on time-domain analyses. To this end, we first performed PSD analysis on highly sampled ECG signals that were highpass filtered with a cutoff frequency of 150 Hz. Our PSD analysis showed that most of the spectral power is largely presented in the low-frequency band between 150–500 Hz. We found that the power in the frequency band between 150 Hz and 1,000 Hz significantly increased during the SNS tasks compared to the baseline for both ECG channels. In particular, the frequency band between 150–250 Hz accounts for the majority of the frequency power (around 60% of the total power). This band, also known as high frequency QRS in studies [43], [44], has been used as an indicator to detect patients with coronary artery disease, which has been noted to be associated with sympathetic overactivity [5]. This suggests that TVSKNA reconstructed using this band (e.g., TVSKNA1) may be able to assess coronary artery disease.

It is interesting to note that our PSD analysis showed inconsistent results. For example, in Table III, with Channel 1 VM and ST, the spectral power change was greater in the second (250–500 Hz) and first frequency bands (150–250 Hz), respectively, when compared to baseline values. In other words, Channel 1 VM showed a decreased mean normalized power change from 63.4% to 57.7% in the first frequency band (150–250 Hz), whereas that of the second frequency band (250–500 Hz) increased from 36.3% to 41.8%. Conversely, *vice versa* for Channel 1 ST. As the absolute power of both first and second frequency bands increased significantly, we can infer that the spectral power increased more in different frequency bands depending on SNS tasks.

Based on our PSD analysis, we proposed a new index of sympathetic tone, which we termed time-varying skin nerve activity (TVSKNA). To compute the index, we decomposed ECG signals using VFCDM into various frequency bands, then reconstructed the SNS-associated signal by summing the amplitude values in the frequency range 160–1,120 Hz. We

TABLE IV
SENSITIVITY ANALYSIS OF SKNA INDICES

| Channel 1 | | VM | | | ST | | | TG (CSP-) | | | TG (CSP+) | | |
|---|---|---|---|---|---|---|---|---|---|---|---|---|---|
| | | *J* | BACC | AUC | *J* | BACC | AUC | *J* | BACC | AUC | *J* | BACC | AUC |
| iSKNA | Max | **1** | **1** | **1** | 0.5 | 0.75 | 0.7 | **1** | **1** | **1** | **1** | **1** | **1** |
| | Mean | 0.64 | 0.82 | 0.9 | 0.5 | 0.75 | 0.72 | 0.93 | 0.97 | 0.99 | 0.95 | 0.97 | 1 |
| | S.D. | 0.98 | 0.99 | 1 | 0.43 | 0.71 | 0.7 | 0.93 | 0.97 | 0.99 | **1** | **1** | **1** |
| TVSKNA1 | Max | 0.96 | 0.98 | 0.96 | 0.43 | 0.71 | 0.77 | **1** | **1** | **1** | **1** | **1** | **1** |
| | Mean | 0.81 | 0.9 | 0.92 | **0.86** | **0.93** | **0.95** | **1** | **1** | **1** | **1** | **1** | **1** |
| 160-1,120 Hz | S.D. | 0.96 | 0.98 | 0.96 | 0.36 | 0.68 | 0.67 | **1** | **1** | **1** | **1** | **1** | **1** |
| TVSKNA2 | Max | 0.96 | 0.98 | 0.96 | 0.5 | 0.75 | 0.74 | **1** | **1** | **1** | **1** | **1** | **1** |
| | Mean | 0.78 | 0.89 | 0.92 | 0.64 | 0.82 | 0.87 | **1** | **1** | **1** | **1** | **1** | **1** |
| 320-1,120 Hz | S.D. | 0.96 | 0.98 | 0.96 | 0.29 | 0.64 | 0.59 | **1** | **1** | **1** | **1** | **1** | **1** |
| TVSKNA3 | Max | 0.94 | 0.97 | 0.96 | 0.64 | 0.82 | 0.8 | **1** | **1** | **1** | **1** | **1** | **1** |
| | Mean | 0.89 | 0.95 | 0.95 | 0.71 | 0.86 | 0.9 | **1** | **1** | **1** | **1** | **1** | **1** |
| 480-1,120 Hz | S.D. | 0.94 | 0.97 | 0.96 | 0.57 | 0.79 | 0.76 | **1** | **1** | **1** | **1** | **1** | **1** |
| Channel 2 | | VM | | | ST | | | TG (CSP-) | | | TG (CSP+) | | |
| | | *J* | BACC | AUC | *J* | BACC | AUC | *J* | BACC | AUC | *J* | BACC | AUC |
| iSKNA | Max | 0.92 | 0.96 | 0.99 | 0.29 | 0.64 | 0.63 | 0.58 | 0.79 | 0.76 | 0.77 | 0.89 | 0.91 |
| | Mean | 0.68 | 0.84 | 0.88 | 0.29 | 0.64 | 0.6 | 0.56 | 0.78 | 0.8 | 0.64 | 0.82 | 0.88 |
| | S.D. | 0.96 | 0.98 | 1 | 0.36 | 0.68 | 0.6 | 0.59 | 0.79 | 0.83 | 0.77 | 0.89 | 0.93 |
| TVSKNA1 | Max | 0.98 | 0.99 | 1 | 0.29 | 0.64 | 0.61 | **0.92** | **0.96** | **0.99** | **0.91** | **0.95** | **0.98** |
| | Mean | 0.92 | 0.96 | 0.99 | **0.36** | **0.68** | **0.67** | **0.92** | **0.96** | **0.99** | **0.91** | **0.95** | **0.98** |
| 160-1,120 Hz | S.D. | 0.98 | 0.99 | 1 | 0.21 | 0.61 | 0.53 | 0.69 | 0.85 | 0.87 | 0.68 | 0.84 | 0.91 |
| TVSKNA2 | Max | **1** | **1** | **1** | 0.29 | 0.64 | 0.57 | 0.89 | 0.94 | 0.98 | 0.83 | 0.92 | 0.97 |
| | Mean | 0.94 | 0.97 | 0.99 | 0.21 | 0.61 | 0.57 | **0.92** | **0.96** | **0.99** | 0.85 | 0.92 | 0.98 |
| 320-1,120 Hz | S.D. | 0.98 | 0.99 | 1 | 0.29 | 0.64 | 0.47 | 0.78 | 0.89 | 0.95 | 0.83 | 0.92 | 0.93 |
| TVSKNA3 | Max | **1** | **1** | **1** | 0.21 | 0.61 | 0.55 | 0.85 | 0.92 | 0.93 | 0.76 | 0.88 | 0.93 |
| | Mean | 0.93 | 0.97 | 0.99 | 0.29 | 0.64 | 0.63 | 0.85 | 0.92 | 0.97 | 0.79 | 0.89 | 0.95 |
| 480-1,120 Hz | S.D. | **1** | **1** | **1** | 0.29 | 0.64 | 0.5 | 0.92 | 0.96 | 0.97 | 0.74 | 0.87 | 0.92 |

* $p$<.05, ** $p$<.001. Bold fonts indicate the highest Youden's J (J), balanced accuracy (BACC), and AUC for each SNS stimulation task.

then compared our indices with iSKNA, derived from ECG bandpass-filtered between 500–1,000 Hz, which is the most commonly used frequency band. All SKNA indices, including those extracted from iSKNA and TVSKNA, exhibited significantly higher values during VM, ST, and TG compared to the baseline, except for ST Channel 2. However, our TVSKNA indices showed generally higher sensitivity compared to iSKNA. Moreover, our TVSKNA indices showed more consistency across subjects with generally lower coefficient of variance compared to iSKNA.

Additionally, three different frequency bands were tested using TVSKNA: 160–1,120 Hz (TVSKNA1), 320–1,120 Hz (TVSKNA2), and 480–1,120 Hz (TVSKNA3). TVSKNA1 includes around 99% of the total spectral power. We also tested TVSKNA2 and TVSKNA3 using the same frequency band as TVSKNA1 without 150–250 and 150–500 Hz, respectively. TVSKNA1 showed the highest sensitivity in response to ST and TG stimuli compared to TVSKNA2 and TVSKNA3. For VM, all TVSKNA indices showed comparable sensitivity. Overall, the three TVSKNA frequency ranges exhibited high sensitivity, indicating that even when SKNA is highpass-filtered at a higher cutoff frequency to mitigate EMG artifacts, TVSKNA indices remain more reliable and sensitive compared to iSKNA indices.

Traditionally, electrodermal activity (EDA) and heart rate variability (HRV) have been widely used to assess SNS activity. A study proposed a time-varying EDA index based on VFCDM, called TVSymp, which has shown high sensitivity in detecting SNS activity compared to HRV [45]. Additionally, TVSymp has demonstrated high sensitivity compared to not only time domain but also spectral indices, including EDASymp (spectral power between 0.045–0.25 Hz) and its normalized index, $EDASymp_n$ [28], [45], in detecting

TABLE V
AVERAGE OF COEFFICIENT OF VARIANCE BETWEEN THE BASELINE AND EACH SNS TASK AND INTRA-CLASS CORRELATION RESULTS

| Channel 1 | | VM | | ST | | TG (CSP-) | | TG (CSP+) | |
|---|---|---|---|---|---|---|---|---|---|
| | | CV | ICC | CV | ICC | CV | ICC | CV | ICC |
| iSKNA | Max | 0.28 | 0.99 | 0.61 | 0.67 | 0.42 | 0.99 | 0.64 | 0.98 |
| | Mean | 0.28 | 0.96 | 0.19 | 0.8 | 0.33 | 0.99 | 0.34 | 0.98 |
| | S.D. | 0.32 | 0.99 | 0.39 | 0.32 | 0.52 | 0.98 | 0.43 | 0.99 |
| TVSKNA1 | Max | 0.31 | 0.98 | 0.44 | 0.8 | 0.19 | 1 | 0.34 | 1 |
| | Mean | **0.25** | 0.94 | **0.1** | 0.96 | 0.13 | 1 | **0.17** | 1 |
| 160-1,120 Hz | S.D. | 0.39 | 0.98 | 0.39 | 0.07 | 0.44 | 0.99 | 0.38 | 0.99 |
| TVSKNA2 | Max | 0.34 | 0.98 | 0.4 | 0.78 | 0.23 | 1 | 0.38 | 1 |
| | Mean | **0.25** | 0.94 | 0.11 | 0.91 | **0.11** | 1 | 0.18 | 1 |
| 320-1,120 Hz | S.D. | 0.36 | 0.98 | 0.36 | 0 | 0.42 | 0.99 | 0.39 | 0.99 |
| TVSKNA3 | Max | 0.38 | 0.98 | 0.43 | 0.8 | 0.21 | 1 | 0.62 | 0.99 |
| | Mean | 0.27 | 0.95 | 0.12 | 0.92 | 0.37 | 0.99 | 0.37 | 1 |
| 480-1,120 Hz | S.D. | 0.35 | 0.98 | 0.28 | 0.53 | 0.21 | 1 | 0.5 | 0.99 |
| Channel 2 | | VM | | ST | | TG (CSP-) | | TG (CSP+) | |
| | | CV | ICC | CV | ICC | CV | ICC | CV | ICC |
| iSKNA | Max | 0.4 | 0.98 | 0.73 | 0.48 | 0.36 | 0.9 | 0.8 | 0.92 |
| | Mean | 0.42 | 0.92 | 0.34 | 0.25 | 0.33 | 0.93 | 0.3 | 0.94 |
| | S.D. | 0.38 | 0.98 | 0.55 | 0 | 0.5 | 0.92 | 0.64 | 0.94 |
| TVSKNA1 | Max | 0.31 | 0.99 | 0.34 | 0.18 | 0.21 | 0.98 | 0.62 | 0.96 |
| | Mean | **0.14** | 0.99 | **0.14** | 0.72 | **0.15** | 0.99 | **0.19** | 0.99 |
| 160-1,120 Hz | S.D. | 0.31 | 0.99 | 0.24 | 0 | 0.44 | 0.93 | 0.53 | 0.95 |
| TVSKNA2 | Max | 0.24 | 0.99 | 0.5 | 0 | 0.3 | 0.96 | 0.55 | 0.96 |
| | Mean | **0.14** | 0.99 | 0.15 | 0.21 | 0.16 | 0.98 | 0.21 | 0.99 |
| 320-1,120 Hz | S.D. | 0.28 | 0.99 | 0.32 | 0 | 0.54 | 0.92 | 0.54 | 0.96 |
| TVSKNA3 | Max | 0.25 | 1 | 0.53 | 0.16 | 0.4 | 0.95 | 0.58 | 0.98 |
| | Mean | 0.17 | 0.99 | **0.14** | 0.61 | 0.42 | 0.96 | 0.38 | 0.99 |
| 480-1,120 Hz | S.D. | 0.27 | 0.99 | 0.3 | 0 | 0.35 | 0.97 | 0.6 | 0.98 |

Bold font indicates the lowest coefficient of variance (CV). ICC: intra-correlation coefficients stress, pain, and cognitive changes [29], [30], [34], [46]. We also found that TVSKNA showed higher sensitivity compared to the spectral power of SKNA in the similar frequency range (Table VI). This may be because PSD does not account for time-varying neural activities of SKNA, hence, it is not a good approach to assess SKNA. Recently, some studies demonstrated that SKNA is more sensitive than HRV and EDA indices are in response to SNS stimuli [18], [19]. In our study, TVSKNA indices, including maximum, mean, and standard deviation, outperformed iSKNA indices in terms of sensitivity and reliability. Therefore, our indices are likely more accurate than EDA and HRV indices. Recently, Xing and colleagues, some of the early adopters of SKNA, proposed an iSKNA feature called SKNA energy ratio (SKNAER), based on the ratio between bursts and non-bursts of SKNA signals. This feature showed higher discriminatory power between baseline and SNS activity compared to aSKNA and vSKNA [19]. However, this requires an accurate detection of bursts, and their method was not effective on our data. Another study showed a promising result for burst detection on iSKNA data [47]. Accurate burst detection on TVSKNA data should be addressed in future studies.

TABLE VI

COMPARISON OF AUCS BETWEEN PSD AND TVSKNA, WHICH WERE OBTAINED BETWEEN THE BASELINE AND EACH SNS TASK.

| Channel 1 | Normalization | VM | ST | TG (CSP-) | TG (CSP+) |
|---|---|---|---|---|---|
| Absolute PSD Power (150 – 1,000 Hz) | | 0.88 | 0.69 | 0.99 | **1** |
| | X | 0.76 | 0.49 | 0.68 | 0.63 |
| TVSKNA1 (160 –1,120 Hz) | Max | **0.96** | 0.77 | **1** | **1** |
| | Mean | 0.92 | **0.95** | **1** | **1** |
| | S.D. | 0.96 | 0.67 | **1** | **1** |
| **Channel 2** | **Normalization** | **VM** | **ST** | **TG (CSP-)** | **TG (CSP+)** |
| Absolute PSD Power (150 – 1,000 Hz) | | 0.96 | 0.58 | 0.82 | 0.89 |
| | X | 0.67 | 0.51 | 0.65 | 0.71 |
| TVSKNA1 (160 –1,120 Hz) | Max | **1** | 0.61 | **0.99** | **0.98** |
| | Mean | 0.99 | **0.67** | **0.99** | **0.98** |
| | S.D. | **1** | 0.53 | 0.87 | 0.91 |

Bold fonts indicate the highest AUC for each SNS stimulation task. Normalization was done by dividing absolute PSD power by total power.

Regarding electrode locations, both channels showed differences in PSD and sensitivity. One study demonstrated differences in performance even for the same lead, depending on the location (chest vs. biceps vs. forearm) [19]. In our study, Channel 1 (ECG lead 1) exhibited a larger range of increased absolute power compared to Channel 2 (ECG lead 3). Moreover, Channel 1 showed higher sensitivity in all indices compared to Channel 2. In particular, the CSP- condition exhibited generally lower sensitivity from iSKNA indices in Channel 2 (~0.6 of Youden's J), while SKNA indices from Channel 1 showed higher sensitivity (~1 of Youden’s J). On the other hand, TVSKNA indices demonstrated higher sensitivity in both channels for CSP-.

TVSKNA showed higher sensitivity than iSKNA did, while having a lower coefficient of variance across subjects. Therefore, TVSKNA can improve current SKNA-based analyses for many affective research problems, including detection, quantification, and analysis of affective states (emotion, stress, etc.). However, our study has a few limitations. First, we manually identified and filtered noise components which were presented in both the baseline and the SNS tasks, which is one of the major issues when collecting data at high sampling frequencies (e.g., 4,000+ Hz). The recording device can be impacted by nearby electronic devices. Note that we removed noise frequencies only when both the baseline and the SNS tasks showed the same noise component concurrently. As this is an impractical approach, future studies need to develop an automatic algorithm to detect and remove noise frequencies, such as adaptive filtering. Second, both motion and muscle artifact noise may be present in SKNA signals, as ECG recordings are also susceptible to EMG artifacts (which primarily contain energy up to 400 Hz) [15]. To mitigate this interference, iSKNA is typically processed using a bandpass filter with cutoff frequencies of 500–1,000 Hz [15]. However, a recent study found that muscle noise can still interfere with SKNA signals within this frequency band [21]. Therefore, simple filtering methods alone do not guarantee noise-free SKNA signals suitable for analysis. Future studies should focus on developing algorithms capable of automatically detecting and removing both motion and muscle artifacts from SKNA recordings. Third, ST in Channel 2 showed no significant increase in absolute power in the frequency domain compared to the baseline. Moreover, none of the SKNA indices obtained during ST Channel 2 showed a significant increase compared to the baseline. ST exhibited generally lower ICC values compared to the other SNS tasks, indicating poorer reliability. In particular, most indices during ST in Channel 2 showed poor reliability across subjects (ICC < 0.5). That is likely because ST involves non-specific stimulation, unlike VM and TG tasks. Furthermore, ST did not show as high sensitivity as the other SNS tasks, which may be because ST does not elicit sufficient stress. Most participants had undergone ST through other experimental protocols (i.e., they were familiar with ST), hence, ST may not have induced the expected level of stress. Additionally, participants were asked to subvocalize the color during the task instead of speaking out loud, which may have been less stressful. Finally, TVSKNA was developed and evaluated using data from eight male and eight female participants, which may limit its generalizability and statistical power. The performance of TVSKNA may vary across populations. However, we hypothesize that both iSKNA and TVSKNA will exhibit similar patterns of performance variation across populations. This should be further investigated in future studies.

## V. CONCLUSIONS

Our PSD analysis showed that the spectral power is largely present in the low frequency band (150–500 Hz). Additionally, power in the frequency band 150–1,000 Hz significantly increased in response to SNS stimuli. The TVSKNA index is a highly sensitive discriminator of the SNS activity induced by pain and stress. Our time-varying spectral index outperformed iSKNA, which has been the most widely used time-series index of SKNA in previous studies and has been demonstrated to be more sensitive than HRV and EDA to SNS response. Our high-performance index showed the feasibility of solving numerous challenges in the field of affective research, including the quantitative assessment of emotion, stress, and other affective states. Despite our successful outcomes, future studies need to address our limitations, discussed above, in order to extend our methods to various SNS-related applications.

ACKNOWLEDGEMENT

This work was supported by the Office of Naval Research N00014-21-1-2255. The data that support the findings of this study are available from the corresponding author upon reasonable request.